\documentclass[%
 reprint,
nofootinbib,
 amsmath,amssymb,
 aps,
 twocolumn
]{revtex4}

\usepackage{amssymb,amsmath,amsfonts,amsbsy,graphicx}
\usepackage{color}
\usepackage{psfrag}
\usepackage{stackrel}

\newcommand\be{\begin{equation}}
\newcommand\ee{\end{equation}}
\newcommand\ba{\begin{eqnarray}}
\newcommand\ea{\end{eqnarray}}\newcommand\eq{\begin{equation}}           
\newcommand\en{\end{equation}}

 \newcommand\lcdm{$\Lambda$CDM }
 
 \newcount\colveccount
\newcommand*\colvec[1]{
        \global\colveccount#1
        \begin{pmatrix}
        \colvecnext
}
\def\colvecnext#1{
        #1
        \global\advance\colveccount-1
        \ifnum\colveccount>0
                \\
                \expandafter\colvecnext
        \else
                \end{pmatrix}
        \fi
}
\input epsf

\usepackage{ulem}
\usepackage{amssymb}
\usepackage{amsmath}
\usepackage{bm}
\usepackage{graphicx}
\usepackage{color}
\usepackage{subfig}
\usepackage{caption}

\def\gsim{\;\rlap{\lower 2.5pt
 \hbox{$\sim$}}\raise 1.5pt\hbox{$>$}\;}
\def\lsim{\;\rlap{\lower 2.5pt
 \hbox{$\sim$}}\raise 1.5pt\hbox{$<$}\;}

\usepackage{graphicx}
\usepackage{url}
\usepackage{amssymb,amsmath}
\usepackage{amsbsy}

\usepackage{graphicx}
\usepackage{url}
\usepackage{amssymb,amsmath}
\usepackage{amsbsy}

\begin{document}
\title{
Boosting small-scale structure via primordial black holes and implications for sub-GeV dark matter annihilation
}
\author{Kenji Kadota$^1$ and Joseph Silk$^{2,3,4}$ \\
  {\small $^1$ Center for Theoretical Physics of the Universe, Institute for Basic Science (IBS), Daejeon, 34051, Korea} \\
  {\small $^2$ Institut d’Astrophysique de Paris, UMR 7095 CNRS, Sorbonne Universit\'{e}s, 98 bis, boulevard Arago, F-75014, Paris, France} \\
    {\small $^3$The Johns Hopkins University, Department of Physics and Astronomy, Baltimore, Maryland 21218, USA
    }
    \\
     {\small  $^4$ Beecroft Institute for Particle Astrophysics and Cosmology, University of Oxford, Oxford OX1 3RH, UK
} 
}

\begin{abstract}
  We explore the possibility that the annihilation of dark matter (DM) is boosted due to enhanced substructure in the presence of  primordial black holes (PBHs) which constitute a sub-component of DM. The PBHs can generate  entropy fluctuations at the small scales which  trigger  early structure formation, and a large fraction of the whole DM can reside in these collapsed objects that formed at high redshift ($z\gtrsim 100$). Such early forming minihalos consequently possess  higher densities than those in the conventional scenarios (without PBHs) and would be more resilient to  tidal disruptions. Our scenarios of the annihilation boost due to DM substructures are of particular interest for light ($< 1$ GeV) DM which has been less explored compared to heavier DM in the presence of the PBHs.

\end{abstract}

\maketitle   

\setcounter{footnote}{0} 
\setcounter{page}{1}\setcounter{section}{0} \setcounter{subsection}{0}
\setcounter{subsubsection}{0}

\section{Introduction}
The nature of dark matter (DM) still remains an open question and there are a wide range of possibilities, such as WIMPs (weakly interacting massive particles) and axions \cite{Jungman:1995df,Bertone:2004pz,Peccei:1977hh,Weinberg:1977ma,Wilczek:1977pj,Preskill:1982cy,Abbott:1982af,Dine:1982ah}.  Possible DM candidates  are not limited to fundamental particles. Gravitational wave detections from black hole binaries have revived  interest in primordial black holes (PBHs) as a DM candidate, even though accumulating  data from gravitational microlensing and other astrophysical constraints has been narrowing  possibilities for PBHs to constitute all of  the  DM \cite{Carr:2020xqk,Green:2020jor,Carr:2020gox}. PBHs as a partial DM component are however still an intriguing possibility, and we consider here scenarios where self-annihilating DM  is the dominant component of  DM while the PBHs constitute a sub-dominant component.

Our scenarios focus on the effects of PBHs whose random distributions introduce Poisson noise in the density, which can amplify the small-scale matter power spectrum. This results in  early structure formation (at redshift $z>100$) where more than half of the total DM can  be in  collapsed halos by $z=100$. This is in stark contrast to  conventional \lcdm without PBHs for which the typical halo collapse epochs are below $z=20$. A characteristic feature of those early-forming minihalos is the high density (dependent on the formation redshift $\propto (1+z)^3$) which would help minihalos survive  tidal disruption on infall into larger structures. The astrophysical probes on small-scale structures in the presence of Poisson fluctuations have been widely studied, including  the Ly-$\alpha$ forest, gravitational lensing and 21cm observations \cite{Afshordi:2003zb,Inman:2019wvr,Murgia:2019duy,Oguri:2020ldf,Gong:2017sie,Carr:2018rid,Mena:2019nhm}.  
We discuss the DM annihilation boost (by a factor $10^3\sim 10^8$ depending on the PBH parameters) due to the enhanced substructures and we also briefly discuss their possible probes by x-ray/gamma-ray experiments. The PBH Poisson fluctuation effects discussed in this paper are based on structure formation through gravity and do not depend on the microscopic properties of DM. We hence first give a generic discussion of matter fluctuation evolution without specifying  DM properties such as particle masses and interactions. In discussing  DM annihilations in a later section, we specify their properties, 
treating the DM mass and annihilation 
cross-section as  free parameters.

We note however  that our studies should be of particular interest for  light ($<1$ GeV) DM which has not yet been fully  explored  in the presence of PBHs. Mixed DM scenarios consisting of annihilating DM and PBHs have indeed been discussed along with DM halo formation around the PBHs, and the incompatibility of thermal WIMPs and PBHs has also been pointed out \cite{Gondolo:1999ef,Lacki:2010zf,Boucenna:2017ghj,Adamek:2019gns,Eroshenko:2016yve,Carr:2020mqm,Cai:2020fnq,Delos:2018ueo,Kohri:2014lza,Bertone:2019vsk}. The effects due to  DM accretion onto the PBHs studied in the previous literature are independent of the Poisson effects to be discussed here.  DM can accrete onto the PBH and form a steep DM density profile around the PBH, which results in  enhanced annihilation from the dense inner region of such a compact DM halo. The non-observation of such annihilation signals results in tight bounds on model parameters, excluding the thermal WIMPs in the presence of  PBHs.  Stringent bounds from DM accretion arise especially when the thermal kinetic energy of accreting DM is negligible compared with its potential energy, so that the kinetic energy does not impact the DM density profile. On the other hand, the ratio of the DM kinetic energy to the potential energy becomes bigger for a lighter PBH mass as well as for lighter DM. To avoid these potentially tight bounds from DM accretion, a light PBH mass (sub-solar mass range) is also of particular interest in our study (hence the black holes are necessarily primordial). 
Previous studies on  DM matter accretion have however  investigated only a  limited DM/PBH parameter space and mainly focused on the heavy ($m_{\chi}>1$ GeV) DM mass (often with the fixed annihilation cross-section $\langle \sigma v\rangle\sim 3\times 10^{-26}cm^3/s$).
The incompatibility of the mixed WIMP/PBH scenarios hence cannot be simply applicable once one extends the available parameter space, and it is of considerable interest to investigate the expanded parameter space beyond the simple WIMP paradigms. A small portion of the DM consisting of  PBHs could well help with  DM detection in an  unexpected region of DM parameter space through  enhanced annihilation signals.

Our paper is structured as follows. \S \ref{density} reviews the matter fluctuations in the presence of the PBHs along with comparison to  standard \lcdm cosmology. After describing the enhanced formation of the 
small-scale structures, \S \ref{survive} discusses the possibilities for these dense minihalos to survive  tidal disruptions. As an interesting phenomenology for detecting  any effects of these small-scale structures, \S \ref{DMannihilate} assumes the dominant component of DM can self-annihilate and estimates the DM annihilation boost factor. 
\section{Density fluctuations and  minihalo formation in the presence of  PBHs}
   \label{density}
We assume the PBHs are randomly distributed. This  is a reasonable assumption considering that the typical separation between the PBHs would be larger than the horizon scale at the formation of PBHs, and the PBH contribution to the power spectrum is given by Poisson fluctuations \cite{Afshordi:2003zb,Murgia:2019duy,Inman:2019wvr,Meszaros:1975ef,1977A&A....56..377C,Chisholm:2005vm,Kashlinsky:2016sdv,1984MNRAS.206..801C}
\ba
P_{PBH}(k)=
\frac{1}{n_{PBH}}
\ea
where $n_{PBH}$ is the comoving number density
\ba
n_{PBH}=\frac{\Omega_{DM} \rho_{cri} f_{PBH}}{M_{PBH}}
\ea
$f_{PBH}\equiv \Omega_{PBH}/\Omega_{DM}$. 
The PBH fluctuations are only in the PBH component and are independent of the adiabatic fluctuations, and one can interpret the PBH perturbations as  isocurvature perturbations.
The total power spectrum is the sum of the conventional adiabatic perturbations and the PBH perturbations
\ba
P(k,z)=D^2(z) \left( T_{ad}^2(k)P_{ad}(k) + T^2_{iso}(k) P_{iso}(k)\right)
\ea
where $P_{iso}=f^2_{PBH} P_{PBH}$, $D(z)$ is the linear growth factor normalized by $D(0)=1$ and $T$ is the transfer function. $T_{iso}$ is \cite{1999coph.book.....P}
\ba
T_{iso}(k)=\frac{3}{2} (1+z_{eq}) \mbox{ for } k_*>k>k_{eq}\\
T_{iso}(k)=0 \mbox{ otherwise }
\ea
We conservatively truncate the transfer function for $k>k_*$ corresponding to  scales smaller than the mean separation between PBHs \cite{Inman:2019wvr,Oguri:2020ldf}
\ba
k_*&=& \frac{2\pi}{(3/4\pi n_{PBH})^{1/3}}\\
& \sim &
4\times 10^4 [h/Mpc]
\left( \frac{M_{\odot}/h}{M_{PBH}}\right)^{1/3} \left( \frac{\Omega_m}{0.3}\right)^{1/3} f_{PBH}^{1/3}
\nonumber
\ea
because, for a very small  scale, we cannot apply the linear theory with the ideal fluid approximation and other effects besides Poisson noise (discreteness) fluctuations, such as clustering fluctuations, could be more important \cite{Inman:2019wvr,Desjacques:2018wuu,Ali-Haimoud:2018dau}.



\begin{figure}[!htbp]

   \begin{tabular}{c}

                          \includegraphics[width=0.5\textwidth]{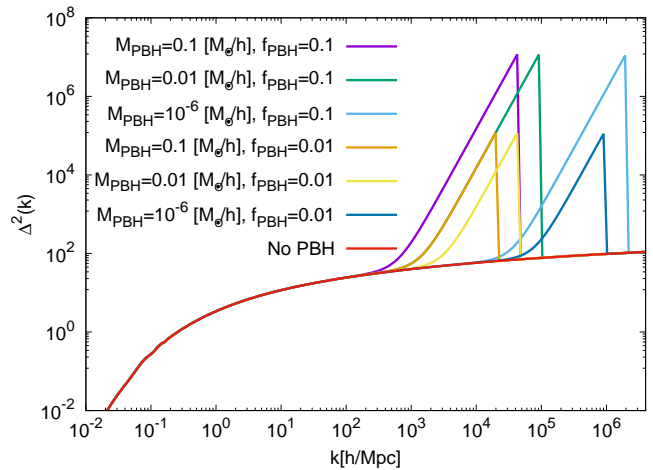} 
    \end{tabular}

    
   \caption{
     The linear matter power spectrum in the presence of PBHs at $z=0$. The power spectrum for the conventional case  \lcdm without PBHs is also shown for comparison. 
   }
%

   \label{kvspkoct19}
\end{figure}
The dimensionless power spectra $\Delta^2(k)=P(k)k^3/2\pi^2$ extrapolated to $z=0$ are shown in Fig. \ref{kvspkoct19}.
\begin{figure}[htbp]
        \includegraphics[width=0.5\textwidth]{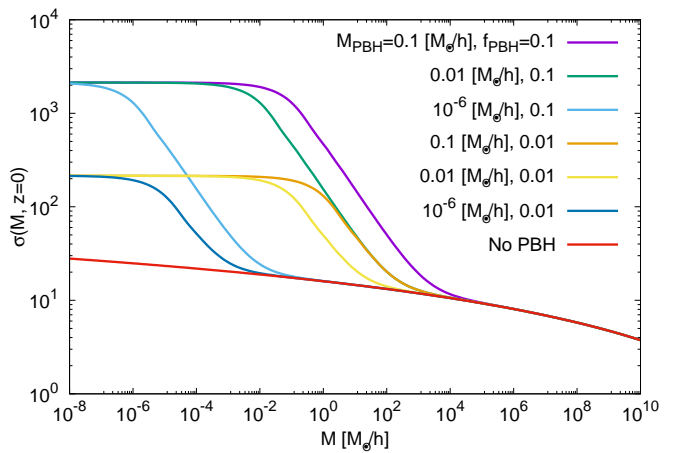} 
   \begin{tabular}{c}
    \end{tabular}

    
   \caption{
    The rms fluctuations $\sigma(M, z=0)$ in the presence of PBHs, to be compared with that for the conventional case \lcdm without PBHs.
   }

%
   
   \label{sigmaz0}
\end{figure}
While there are tight bounds on the perturbation spectrum down to galaxy scales, a non-standard spectrum deviating from a simple power law at smaller scales can still remain an interesting possibility. The linear power spectrum smoothed over the comoving scale $R$ has variance 
\ba
\sigma^2(M,z)&=&\int d\ln k \frac{k^3 P(k,z)}{2\pi^2} \left| W(kR)\right|^2 \\  \nonumber
&=&\int \frac{d\ln k}{2\pi^2} k^3 P(k) T^2(k) D^2(z) \left| W(kR)\right|^2
\ea
where $W(kR)=3\left[\sin(kR)-(kR)\cos(kR)\right]/(kR)^3$ is the Fourier transform of the real-space spherical top-hat window that contains mass $M$. 
 We can see, from Fig. \ref{sigmaz0}, the enhanced variance at small scales, and we expect that small halos can be produced much earlier in the presence of the PBH Poisson fluctuations. We characterize our minihalo formation epochs, according to the spherical collapse model, by the redshift satisfying
\ba
\frac{\delta_c}{\sigma(M,z_c)}=\frac{\delta_c}{\sigma(M,z=0)D(z_c)}=1
\ea
where $\delta_c=1.686$ is the critical overdensity for collapse (we ignore its weak dependence on the redshift and cosmology)
\footnote{The halos for a given mass are formed over a wide range of redshifts, and there is no unique way to define the halo formation epochs. There are a variety of characteristic redshifts used in the literature, such as the 'half-mass' formation redshift of a halo defined as its earliest time when at least half of its total halo mass has been assembled into a single progenitor \cite{Neto:2007vq,Lacey:1993iv,Giocoli:2006yz,Ludlow:2013vxa,Sanchez-Conde:2013yxa, Maccio:2008pcd}. We are on the other hand more interested in the redshift ($z\gtrsim 100$) when the first generation halos formed. These can be dense enough to survive through  tidal disruptions up to the present time.}.
The rarer fluctuations with a larger amplitude $N\sigma$ collapse earlier at a higher redshift $z\sim N z_{c}$ if the density fluctuations have a Gaussian probability distribution. 
 \begin{figure}[htbp!]
             \includegraphics[width=0.5\textwidth]{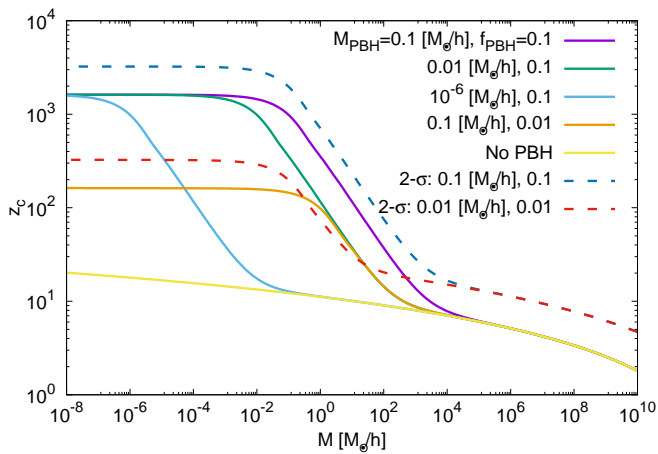} 
   \begin{tabular}{c}

    \end{tabular}

    
   \caption{
     The collapse redshifts for the 1 $\sigma$ fluctuations (solid) as a function of a halo mass. The collapse redshifts for the 2 $\sigma$ fluctuations are also shown as dashed lines.
   }

  
   \label{collapsezoct19}
\end{figure}
 Fig. \ref{collapsezoct19} shows such collapse redshifts for 1$\sigma$ and 2$\sigma$ peaks. The objects collapsed from the bigger peaks are rarer, but they form earlier and hence are more resilient to  tidal disruptions due to the higher density. The final abundance of minihalos at the present epoch would depend on these competing effects.
 Note that the variances and consequently the collapse redshifts share the same flat plateaus for low halo masses among different values of $M_{PBH}$. This is a consequence of the common peak height of $P_{iso}$ among different PBH masses. While the smaller $M_{PBH}$ makes the amplitude of $P_{iso}(k)$ smaller for a given $k$, a smaller $M_{PBH}$ makes the cutoff scale $k_*$ bigger. Our parameterization of the PBH fluctuations hence keeps the same peak amplitude for a different value of $M_{PBH}$. 
  This is in contrast to the effect of $f_{PBH}$ whose smaller value makes both $P_{iso}(k)$ amplitude and $k_*$ smaller.
We also note there is a large slope besides a plateau in Fig. \ref{collapsezoct19}. For the conventional $\Lambda$CDM, the small-scale matter power spectrum has the effective spectral index $n\sim n_s -4 \sim -3$ ($n_s\sim 1$ is the primordial spectral index and $-4$ comes from the transfer function), so that the variance $\sigma(M) \propto M^{-(n+3)/6}$ has a weak dependence on mass scale. This results in the almost simultaneous formation of small structures across a wide range of mass scales, and it is still an open question as to how the structure formation proceeds at high redshift. Using  hierarchical structure formation as a guiding principle could be misleading here \cite{Widrow:2009ru,Berezinsky:2014wya}. For instance, the small collapsed halos may not be fully virialized by the time they are captured by the bigger halos, and small clumps may be more susceptible to tidal destruction because the high core density profile may not be still formed by the capture epochs. 
   The Poisson fluctuations on the other hand possess $n\sim 0$ and there is a steep $k$-dependent feature in the matter power spectrum, and consequently the formation is not as simultaneous as in the \lcdm scenario illustrated as the steep slopes in Fig. \ref{collapsezoct19}. These small clumps at the corresponding scales in the spectrum may form well before the larger halos form, so that they could well survive  tidal disruptions.

\begin{figure}[htbp]

   \begin{tabular}{c}

                 \includegraphics[width=0.5\textwidth]{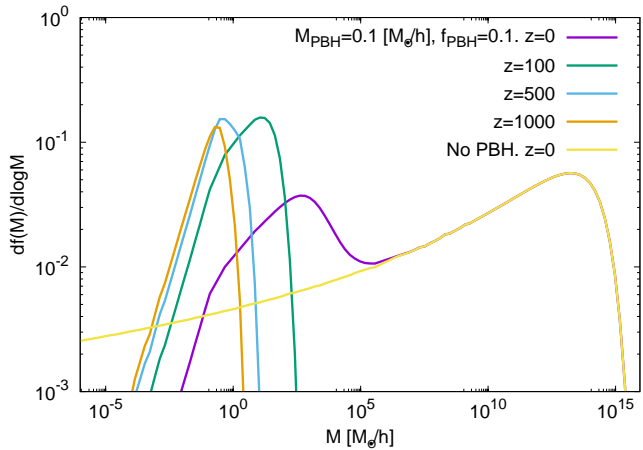} 
    \end{tabular}

    
   \caption{
  The mass fraction of isolated collapsed halos $df(M)/d \log M$ using the Press-Schechter formalism. The mass fractions for the \lcdm scenario without PBHs are too small to be shown in this figure at  high redshifts.
  } 
      \label{massfracz}
\end{figure}

It would also be illustrative to see the mass fraction contained in these collapsed objects at a high redshift in our PBH scenarios. We follow the Press-Schechter formalism where the mass function (the comoving number density of halos), $n(m,z)$, is given by \cite{1974ApJ...187..425P}
\ba
n(m,z)dm =\frac{\bar{\rho}}{m}\frac{d f(m,z)}{dm} dm
\ea
$\bar{\rho}$ is the background density and $f$ represents the fraction of mass which is locked in collapsed halos. We plot the mass fraction $f$ for different redshifts in Fig. \ref{massfracz} (using the mass function in Ref. \cite{1999MNRAS.308..119S} where $\sigma$ is calculated including the PBH isocurvature perturbations), which illustrates a stark contrast between our PBH scenarios and the conventional \lcdm without the PBHs.

We can see a generic trend here in  hierarchical structure formation. The first small halos form through direct collapse of peaks in the density field, and the collapsed fraction for a smaller smoothing mass scale decreases as small halos are assembled to form larger halos (these minihalos become subhalos). The peak of the mass fraction hence shifts toward a higher halo mass at a lower redshift. For the PBH parameters shown in Fig. \ref{massfracz}, the peaks of the matter power spectra show up around $k \sim 10^4$ to $10^{5} h/Mpc$ as shown in Fig. \ref{kvspkoct19}. Such peaks correspond to mass scales of order $0.01\sim 10 M_{\odot}$ and the mass fractions also peak around these mass scales, as expected. The peak positions of the mass fractions are redshift-dependent, and  are characterized by the mass scale satisfying $\delta_c/\sigma(m,z_c)\sim 1$. Heavier halos bigger than such a characteristic mass scale are suppressed due to the form of the mass fraction $\propto e^{-\delta^2_c/2\sigma^2}$. 
The mass fractions for $z\gtrsim 100$ for the standard \lcdm are too small to be shown in this figure.
While the Press-Schechter formalism does not take account of the abundance of subhalos once those small halos merge into larger ones, it can illustrate how those small isolated halos are formed at  high redshifts. This figure hence illustrates the lower bound to the mass fraction in the small dense halos because, in addition to the isolated halos, some of the subhalos can survive even after being assembled into the larger halos.
We  find that about  half of all the DM resides in  collapsed halos by $z=500$ for $f_{PBH}=0.1, \Omega_{PBH}=0.1$ in Fig. \ref{massfracz}. Because the early formed halos have large densities $\propto (1+z_c)^3$ and are hence resilient to  tidal disruptions, these clumps can potentially lead to significant substructure components if they can survive up to the present epoch \cite{vandenBosch:2017ynq,Goerdt:2006hp,Blinov:2019jqc}. 
The exact survival probability of the halos would require high resolution numerical simulations. We would need a dynamical range covering twenty orders of magnitude in the mass over the time scale of $\sim 10$ Gyrs dedicated to the initial conditions with the PBH isocurvature perturbations, and this would be beyond the scope of even the current state-of-the-art simulations \cite{Wang:2019ftp,Ishiyama:2020vao,Sanchez-Conde:2013yxa,Moline:2016pbm,Zhao:2008wd,Prada:2011jf,Ackermann:2015tah}. We however can analytically gain an insight into the resilience of the dense minihalos against tidal disruptions, and in the next section we give arguments to support the survival of  small halos formed at redshifts well exceeding $100$.
\section{Survival of minihalos}
\label{survive}
We here briefly review two main dynamical processes from which  these minihalos can potentially suffer.
The first process we mention is the dynamical friction which causes the minihalos to fall into the host halo's dense central regions where the strong gravitational field may tidally strip the outer parts of minihalos. Let us consider the subhalo of mass $M$ orbiting within the host halo of mass $M_{host}$. 
We can estimate the orbital decay time-scale due to dynamical friction as
\ba
t\sim t_{orbit} \frac{M_{host}}{M} (\ln\Lambda)^{-1} 
\ea
where the orbital time-scale $t_{orbit}\sim R_{vir}/V_{vir} \sim 0.1 t_H$ ($t_H$ is the Hubble time and we used $M_{host}= 200 \rho_{cri} R_{vir}^3 4\pi/3, V_{vir}^2\sim GM_{host}/R_{vir}$ ($\rho_{cri}$ is the critical density)), and the Coulomb logarithm $\ln \Lambda \sim \ln (M_{host}/M)$ \cite{2008gady.book.....B,2010gfe..book.....M}. 
For the minihalos of  interest here (which can even go down to  scales as small as the order of ${\cal O}(10^{-6})M_{\odot}$), the tidal stripping due to the infall into the host halo's central region caused by the dynamical fraction would be negligible thanks to the large mass hierarchy between the minihalo and host halo masses.
For such small minihalos,  tidal disruptions due to stellar encounters can be more important \cite{Zhao:2005mb,vandenBosch:2017ynq,Green:2019zkz,Blinov:2019jqc,Berezinsky:2005py,Berezinsky:2003vn,Arvanitaki:2019rax} \footnote{The minihalos may be disrupted even without dynamical friction if they happen to be in radial trajectories towards the host halo center and they may be also disrupted due to encounters among the minihalos, for which we refer readers to \cite{Berezinsky:2005py,Berezinsky:2003vn,2020AJ....159...49D}. However these processes turn out to give  smaller effects than  minihalo encounters with  stellar objects to be reviewed in this section.}.
Stellar encounters (such as with stars) transfer the energy into a minihalo, and the minihalo's velocity dispersion increases and its binding energy decreases. We refer the readers to other literature for  extensive studies, and here we give a simple estimate based on the impulse approximation (an interaction is treated as instantaneous because a typical time-scale for the minihalo-star interaction is much smaller than the dynamical time-scale of minihalos (orbital time of particles in a minihalo )). When the minihalo of mass $M$ with  size $R$ encounters  a stellar object with  mass $M_*$ and relative velocity $v$ at impact parameter $b$, the change in the velocity of a particle within a minihalo at a position $r$ (relative to the minihalo center) is of order 
\ba
\Delta v \sim \frac{2G  M_* r}{v b^2}
\ea
The total energy input due to the velocity change is given by integrating over the minihalo density $\rho(r)$ in the impulse approximation \cite{2008gady.book.....B}
\ba
\Delta E&=&\frac{1}{2} \int |\Delta v|^2 \rho(r) d^3 r \\  \nonumber
&\sim&  M \left( \frac{ G M_*}{v b^2} \right)^2  \langle r^2\rangle  ,
\langle r^2 \rangle
=
\frac{\int^{R}_{0} d^3 r r^2 \rho(r) }{M}
\ea
One can introduce the characteristic impact parameter $b_*$ inside which a single encounter can entirely disrupt the minihalo by requiring that the energy input is larger than the binding energy $E_b \sim G M^2/R $ 
\ba
b_{*}^2 \sim  \frac{GM_* R_{vir}}{v V_{vir}}
\ea
where we used $R \sim R_{vir}$, $V_{vir}^2 \sim GM/R_{vir}$ and $\langle r^2 \rangle \sim R^2_{vir}$ \cite{Goerdt:2006hp,Tinyakov:2015cgg,Schneider:2010jr}. 
We can then estimate the time-scale for  minihalos orbiting in a host halo to be disrupted by  stellar encounters as 
\ba
t&\sim& \frac{1}{\pi b_{*}^2 \bar{n}_* v}  \nonumber \\
&\sim& 70 Gyr \left( \frac{1+z_{c}}{100} \right)^{3/2}
\left( \frac{10^6 M_{\odot}/kpc^3}{\bar{n}_* M_*} \right)
\ea
where $\bar{n}_*$ is the mean number density of stars in a host halo.
$z_{c}$ represents the epoch when the halo collapses and the decoupling from the Hubble flow occurs, and the redshift dependence comes from the minihalo density at its formation $\rho(z_{c})\propto (1+z_{c})^3$. 
  We hence can infer that the minihalos in a galactic host halo would suffer from  significant disruptions if they formed after $z\sim 100$, even though the inner central regions could well remain intact \cite{Goerdt:2006hp}. On the other hand, our PBH scenario leads to  early structure formation $z> 100$ and those early-formed minihalos can  avoid  disruptions (this is even so if the minihalos are formed from rarer $N(>1)$-$\sigma$ fluctuations typical in the high density regions \cite{Diemand:2005rd,Diemand:2006ey}). 
 We note that  tidal stripping mainly removes the outer parts of the subhalos, and the dense core regions, from where the annihilation signals mostly originate, can  survive.
 Even if the subhalos evolve while losing their mass after accreting onto the host halos, it is still a matter of debate if they completely lose their identities or not \cite{vandenBosch:2017ynq,2015MNRAS.454.1697X,Berezinsky:2007qu,Delos:2019lik,Green:2019zkz,2010PhRvD..81j3529B,Green:2006hh,Goerdt:2006hp,Schneider:2010jr}.
The tightly bound inner cores of minihalos in our scenario can  survive to the present time even though the DM distribution could differ throughout the course of structure formation from that at the formation epoch. 
While the detailed numerical simulations for the survival probability is beyond the scope of even the state-of-the-art simulations, the above simple order-of-magnitude discussions should suffice  to motivate looking into the phenomenology due to  surviving substructures, as discussed  in the next section.

\section{Dark matter annihilation}
\label{DMannihilate}
Our discussions so far outlined the generic features of  structure formation in the presence of  PBHs, which did not necessarily require us to specify the properties of the dominant DM component.
In this section, we turn our attention to a more model-dependent feature of  PBH scenarios, and, as an interesting application of the enhanced substructures, we give an analytical estimation of the DM annihilation boost factor. We here assume that the dominant component of DM can self-annihilate while the other sub-component is comprised of  PBHs. A boost factor gives a ratio of the total annihilation from all the subhalos for a given host halo 
to that from a smooth host halo without assuming any subhalos (without any PBHs either for an easier comparison with the standard \lcdm cosmology studied in the existing literature) \cite{Strigari:2006rd,Kuhlen:2008aw,Moline:2016pbm}. The DM annihilation rate is proportional to the DM density squared, the so-called the $J$ factor $J=\int \rho^2(x) d^3x$. 
For the $J$ factor from a single subhalo, we simply use the NFW-like profile with  scale radius $r_{s}$, scale density $\rho_{s}$, and tidal truncation radius $r_{t}$ above which the subhalo density vanishes
  \ba
  J&=&\int d^3 x  \rho(x)^2  \nonumber \\
  &=& \frac{4\pi}{3}\rho_{s}^2 r_{s}^3 \left( 1-\frac{1}{(1+c_{t})^3}  \right) 
\ea
where $c_{t}\equiv r_t/r_{s}$ and the integration was performed over $r=[0,r_{t}]$ for the NFW profile
  \ba
  \rho(r)=\frac{\rho_{s}}{(r/r_{s})(1+r/r_{s})^2}
  \ea
  We took for concreteness $r_t=0.77 r_{s}$ \cite{Hayashi:2002qv} (the dependence of the boost factor on $r_t$ is weak for $r_t \gtrsim r_s$).
We use the normalization so that the mass of the DM which can annihilate enclosed within $r_{200}$ is $m_{200}\Omega_{\chi}/\Omega_{m}$ ($r_{200}$ is the radius inside which the total enclosed mass is $m_{200}$ with the average matter density $200 \rho_{cri}$)
\ba
\rho_{s}=\frac{\Omega_{\chi}}{\Omega_m} \frac{200 \rho_{cri}}{3}\frac{c^3}{\ln(1+c)-c/(1+c)}
\ea
with concentration parameter $c\equiv r_{200}/r_s$. We also need to know the subhalo abundance for a given host halo. 
 A simple extrapolation of the subhalo mass function, as conventionally done for the standard featureless power spectrum, may not be justifiable for the  peculiar power spectrum discussed in this paper.
It is in fact still a matter of debate how large a  fraction of the total host halo mass can be provided by the subhalos \cite{vandenBosch:2017ynq,Green:2019zkz,Delos:2019tsl,Stoehr:2002ht,Ullio:2002pj,2011MNRAS.410.2309G,2019Galax...7...81Z,Wang:2019ftp,Anderhalden:2013wd}. Here, we  take a simple approach following \cite{Erickcek:2015jza,Blanco:2019eij,Arvanitaki:2019rax} that the minihalos produced at a high enough redshift $z_c(\gtrsim {\cal O}(100))$ can survive till the present epoch and the minihalos follow the underlying DM distribution as the larger halos form. The subhalo mass fraction in a host halo at the present epoch originating from these early forming minihalos can  be estimated from the mass fraction using the Press-Schechter formalism at $z=z_c$.
The corresponding boost factor can hence  be estimated as 
\ba
B= \frac{\sum_i \int d^3x \rho^2 _{i}(x)}{\int d^3 x {\rho}_{mat}^2(x)}
\ea
where $\rho_{mat}$ is the underlying matter density of a host halo (assuming no substructures) and the contribution by summing subhalo $J$ factors in a host halo is obtained as \cite{Cooray:2002dia,Bergstrom:2001jj,Ullio:2002pj,Fornasa:2015qua,SanchezConde:2011ap,Ng:2013xha,Namjoo:2018oyn,Erickcek:2015jza,Blanco:2019eij}
\ba
\label{shboost}
&& \sum_i \int d^3x \rho^2 _{i}(x)=  \\
&&\int_{m_{min}} dm \int_{0}^{R_{200}} 4\pi R^2dR   \left.  J \frac{  \rho_{mat}(R)}{m}\frac{df}{dm} \right\vert_{z=z_c}   \nonumber
\ea
The capital (small) letters refer to the host halo (subhalo). For instance $m$ is the subhalo mass, $R$ is the distance from the host halo center and $R_{200}$ is the host halo radius. 
The spatial distribution of the subhalos inside a host halo is also a topic under an active investigation subject to  resolution issues and possibly to baryonic effects \cite{Han:2015pua,Springel:2008cc,Stref:2016uzb,Kelley:2018pdy,Hayashi:2002qv,Wetzel:2016wro}. Even if the subhalo spatial distributions initially follow the underlying host halo matter density, their distributions can be dynamically changed by the tidal forces. The subhalos of our interest in the PBH scenarios are highly bound objects and their spatial distributions could be different from those in the conventional \lcdm scenario where most of the survived subhalos reside in the outer part of the host halos. We, as customarily done for easier comparison with other literature, simply assume a homogeneous distribution of subhalos, and we leave the spatial dependence of the subhalos such as the concentration parameters as a function of the positions in our PBH scenarios for  future work \footnote{Incorporating some modelings of the subhalo distribution inside a host halo can affect the boost factor estimates by less than 10$\%$ \cite{Moline:2016pbm}.}. 
The simple estimation given by Eq. (\ref{shboost}) is $z_c$-dependent. While one should consider all  redshifts in order to estimate the boost factor, we follow the approach of \cite{Erickcek:2015jza,Blanco:2019eij} and consider the single redshift $z_c$ which gives the maximum contribution to the boost factor (this approach turns out to give an order-of-magnitude match to a more detailed analysis including the effects of merging among the minihalos \cite{Delos:2019dyh}). Noting $J \propto m (1+z)^3$, such an optimal $z_c$ corresponds to the redshift at which $(1+z)^3 \int dm df/dm $ becomes maximum. Note that this approximation does not take account of the annihilation boost contributions from the minihalos formed after $z_c$ or the subhalos inside the minihalos at $z_c$. The actual boost factor hence can be bigger than our simple estimates here if the minihalos present at $z=z_c$ survive up to the present epoch. We only discuss the PBH parameters which lead to the optimal $z_c>100$ in our discussions, because we focus on highly bound minihalos produced at sufficiently high redshifts so that they are resilient to tidal disruptions. For instance, the optimal $z_c\sim 300$ for $M_{PBH}=0.1 M_{\odot},f_{PBH}=0.01$.
We also note that, in performing the subhalo mass integration, we introduced the conventional reference value of the minimal halo mass $m_{min}=10^{-6} M_{\odot}$ for concreteness \cite{Bertschinger:2006nq,Green:2003un,Green:2005fa,Loeb:2005pm,Gondolo:2012vh,Profumo:2006bv, Gondolo:2016mrz,Bringmann:2006mu}.



\begin{figure}

     \begin{tabular}{c}
                 \includegraphics[width=0.5\textwidth]{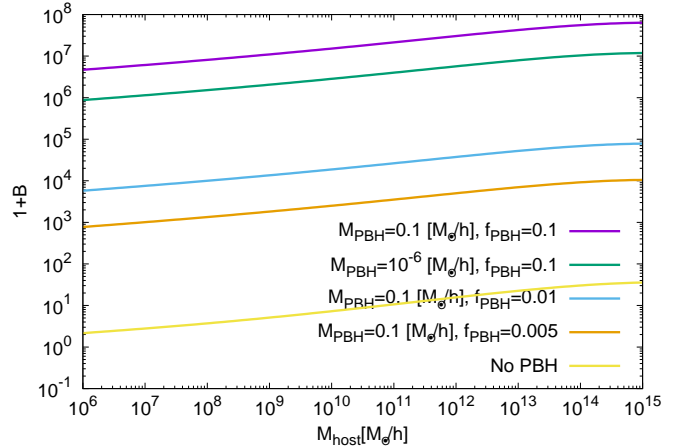} 
    \end{tabular}

    
   \caption{The annihilation enhancement factor, 1+B, as a function of a host halo mass.
   }

   \label{boostvdmhost}
\end{figure}

We assume the NFW profile for a host halo, and our estimated boost factor as a function of a host halo mass is shown in Fig. \ref{boostvdmhost}.\footnote{We follow the convention that the subhalo mass fraction is not subtracted from a host halo (subtraction increases our boost factor estimation) \cite{Moline:2016pbm}.}
For the minihalo concentration parameters at $z=z_c$, we conservatively use the concentration parameter $c=2$ for all the minihalos because it is the minimum concentration parameter value in the simulations performed for a wide range of the masses \cite{Wang:2019ftp,Ishiyama:2020vao,Sanchez-Conde:2013yxa,Moline:2016pbm,Zhao:2008wd,Prada:2011jf,Ackermann:2015tah}.
For the concentration parameters of the host halos, we use the $c-M$ scaling relation in Ref. \cite{Sanchez-Conde:2013yxa} tested for the standard \lcdm cosmology. 
 The boost factor for $f_{PBH}=0.1, M_{PBH}=0.1 M_{\odot}$ is of order $10^7$. We also show the boost factor for $f_{PBH}= 0.005$ and $M_{PBH}=0.1 M_{\odot}$.  For $M_{PBH}=0.1 M_{\odot}$, $f_{PBH}\lesssim 0.005$ leads to the optimal redshift $z_c\lesssim 100$. We do not consider such scenarios because the subhalos may significantly suffer from  tidal disruptions due to late formation. We also showed the boost factor when $M_{PBH}=10^{-6}  M_{\odot}$ for $f_{PBH}=0.1$. For $f_{PBH}=0.1$, $M_{PBH}\lesssim 10^{-7}  M_{\odot}$ corresponds to the cut-off scale of $k_*\gtrsim 4 \times 10^6 [h/Mpc]$ whose corresponding mass scale is $\lesssim 10^{-6} M_{\odot}$. 
 Hence structure formation may suffer from suppression due to  DM 
 free-streaming/acoustic damping if the kinetic decoupling happens to occur at these corresponding scales \cite{Bertschinger:2006nq,Loeb:2005pm}.
We can hence infer that the boost factor could change significantly for $M_{PBH}=10^{-6}M_{\odot}$ if we change the cut-off subhalo mass value $m_{min}=10^{-6} M_{\odot}$. Indeed, using $m_{min}=10^{-5}M_{\odot}$ ($m_{min}=10^{-4}M_{\odot}$) instead of $m_{min}=10^{-6}M_{\odot}$ decreases the boost factor by an order (two orders) of magnitude. On the the hand, for the other PBH parameters shown in Fig. \ref{boostvdmhost}, choosing $m_{min}=10^{-8}M_{\odot}$ to $10^{-4} M_{\odot}$ instead of $m_{min}=10^{-6}M_{\odot}$ does not change the boost factor by an order of magnitude, which is reasonable because the peaks of the mass fractions show up at  mass scales larger than the $m_{min}$ values.
The DM kinetic decoupling can be heavily model-dependent, and we leave a more concrete DM model analysis for future work \cite{Loeb:2005pm,Bertschinger:2006nq,Gondolo:2012vh, Profumo:2006bv,Gondolo:2016mrz,Green:2003un,Green:2005fa,Bringmann:2006mu}.
Fig. \ref{boostvdmhost} shows that the dependence of the boost factor on $M_{PBH}$ is small, compared with that on $f_{PBH}$. This stems from the same reasons mentioned in the previous sections. It is because, while the smaller $M_{PBH}$ makes the amplitude of $P_{iso}(k)$ smaller for a given $k$, a smaller $M_{PBH}$ makes the cut-off scale $k_*$ bigger.
 

 We also note  substructures also arise in conventional \lcdm without PBHs, and these substructures can enhance  DM annihilation rates. Their formation epochs are however much later than those discussed in our PBH scenarios and they are less dense and so  susceptible to tidal disruptions.  The boost factors for \lcdm have been studied numerically and we plot the fitting formula given by Ref. \cite{Sanchez-Conde:2013yxa} (which was tested for the halo mass range $10^6 M_{\odot}<M_{host}<10^{16} M_{\odot}$ assuming the minimal subhalo mass $m_{min}=10^{-6}M_{\odot}$) for comparison.
We can indeed see the significant enhancement of boost factors in the PBH scenarios compared with the standard \lcdm cosmology. Even though more rigorous  estimates of the boost factors due to substructures still require detailed numerical simulations, our studies illustrate the potential significance of the PBHs in the search for  DM annihilation signals.

\begin{figure}[htbp]
  
   \begin{tabular}{c}

               \includegraphics[width=0.5\textwidth]{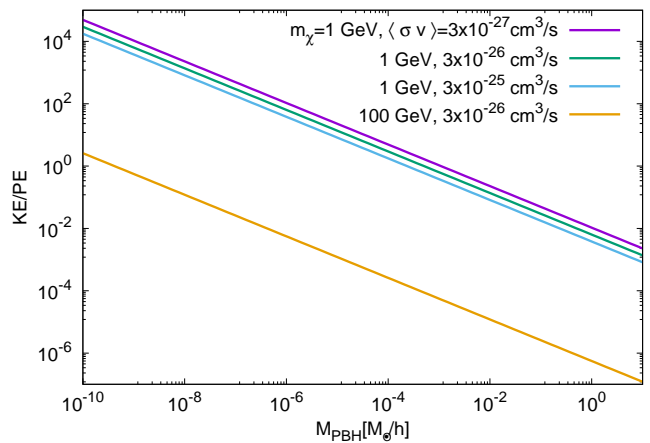}
    \end{tabular}

    
   \caption{
The ratio of the DM thermal kinetic energy over the potential energy at $r_{core}$ as a function of the PBH mass. 
   }
   \label{KEPE}
\end{figure}

In addition to the `Poisson effect' discussed so far where the PBH Poisson fluctuations can collectively trigger the early formation of small-scale structures, there is a `seed effect' where each PBH can individually impact  cosmological structures  \cite{Carr:2018rid,1984MNRAS.206..801C}. For instance, when the PBHs are rare enough such that the interactions between the PBHs can be neglected in the process of DM accretion onto the PBH, the PBH can be a seed for an ultracompact minihalo with a steep DM profile.
Such a dense DM halo around the PBH can also result in DM annihilation enhancement. For instance, if we fix the DM parameters to the canonical WIMP values of $\langle \sigma v \rangle=3 \times 10^{-26}cm^3/s, m_{\chi}=100$ GeV, the lack of any observation of such enhanced annihilation signals from the Fermi Large Area Telescope (LAT) extragalactic gamma-ray observations requires $f_{PBH}\lesssim 2 \times 10^{-9}$ (for $M_{PBH}\gtrsim 10^{-6} M_{\odot}$) \cite{Adamek:2019gns,Ando:2015qda}. 
For such conventional weak-scale WIMP DM parameter values of $\langle \sigma v \rangle$ and $m_{\chi}$, the Poisson fluctuation effects would be small because the isocurvature perturbation amplitude is proportional to $f_{PBH}$. Such a small $f_{PBH}$ however is deduced based on the assumption that the thermal kinetic energy of DM particles is small and does not influence the DM density profile around the PBH \cite{Adamek:2019gns}.
We instead treat the DM parameters as free parameters in this paper for more generic discussions, and we argue in the following that our substructure annihilation boosts accordingly would be of great interest for light (sub-GeV) DM for which the previous bounds on the mixed DM scenarios cannot be simply applied.
We in the following briefly review the discussion of Ref. \cite{Adamek:2019gns}, and refer the readers to existing literature for the derivation of equations and more detailed analysis on the steep DM density profile as well as the consequently enhanced annihilation around the PBHs \cite{Gondolo:1999ef,Lacki:2010zf,Boucenna:2017ghj,Adamek:2019gns,Bertone:2019vsk,Eroshenko:2016yve,Delos:2018ueo,Kohri:2014lza,Carr:2020mqm}.
Following Refs. \cite{Adamek:2019gns,Boucenna:2017ghj} for the spherically symmetric collapse model for the DM halo around the PBH, Fig. \ref{KEPE} shows the ratio of the thermal kinetic energy (KE) and the gravitational potential energy (PE) of a DM particle at a turnaround radius when the DM particle starts moving towards the PBH (Figure (2) of Ref. \cite{Adamek:2019gns} corresponds to $m_{\chi}=100$ GeV and $\langle \sigma v \rangle=3\times 10^{-26}cm^3/s$).
The kinetic energy estimates heavily 
depend on the particle physics model details determining the DM kinetic decoupling epochs, but this simple figure  illustrates the generic features of DM in the presence of the PBHs. The DM potential energy due to the PBH gravitational potential in this figure was estimated as 
$E_p={GM_{PBH}m_{\chi}}/{r_{ta}}$
where, for the turnaround radius $r_{ta}$, $r_{core}$ was used \cite{Adamek:2019gns}
\ba
r_{core}& \approx& 1.2 \times 10^{-6} \left(  \frac{M_{PBH}}{M_{\odot}/h} \right)^{1/3}
\left( \frac{m_{\chi}}{GeV} \right)^{-4/9}
 \nonumber \\
& \times & \left( \frac{\langle \sigma v \rangle /}{3\times 10^{-26} cm^3/s} \right)^{4/9}  [kpc/h]
 \ea
 inside which the DM density is constant (this value is the maximum possible DM density today $\rho_{max}=m_{\chi}/\langle \sigma v \rangle t_{today}$) and beyond which the density possesses a steep profile. 
 To estimate the kinetic energy (or the DM temperature at  turnaround) $E_k\sim T_{\chi}$, one needs to estimate the DM kinetic decoupling temperature. The precise estimation of the kinetic decoupling temperature requires the (possibly momentum dependent) elastic scattering 
 cross-sections with the particles in the plasma, and can be heavily model-dependent \footnote{For instance, see Ref. \cite{Gondolo:2012vh,Bringmann:2006mu} for the detailed estimates of kinetic decoupling temperature.}. To get an order-of-magnitude estimate for this ratio $KE/PE$ and also for an easier comparison with the previous literature, we use a simple dependence of the kinetic decoupling temperature on the DM mass as $T_{KD}/m_{\chi} \propto m_{\chi}^{1/4}$ \cite{Bringmann:2006mu}. In general, the lighter DM kinetically decouples at a lower temperature and it suffers less from the rapid temperature decrease $T_{\chi} \propto a^{-2}$ after the decoupling ($T_{\chi}\propto a^{-1}$ before the decoupling). The ratio KE/PE accordingly increases for a smaller DM mass as shown in Fig. \ref{KEPE}, because the gravitational potential energy decreases and the kinetic energy increases as the DM mass decreases. 
 Fig. \ref{KEPE} illustrates that, for $m_{\chi}\gtrsim 100$ GeV, the thermal kinetic energy is not important at a radius exceeding the core radius for $M_{PBH}\gtrsim 10^{-6} M_{\odot}$. Ref. \cite{Adamek:2019gns} performed the numerical simulations to obtain the bounds $f_{PBH} \lesssim 10^{-9}$ but limited the validity of their simulation results to the parameter ranges where the kinetic energy can be ignored during the DM accretion.
 Hence such quantitative results cannot be simply applied once one extends the available parameter space, for instance, to light ($<1$ GeV) DM and  sub-solar mass PBHs for which Fig. \ref{KEPE} shows the kinetic energy can be significant. Even though the numerical simulations for the DM density profile around the PBH are still yet to be performed when the DM kinetic energy can not be ignored, the analytical studies
 for such cases with non-negligible DM kinetic energy indicate that the accretion bounds on the PBH fraction become weaker when $M_{PBH}$ is smaller \cite{Eroshenko:2016yve,Carr:2020mqm,Cai:2020fnq}.
The discussions for the DM profile around the light (sub-solar mass) PBH for the light DM ($m_{\chi}=1$keV$\sim 1$GeV) are beyond the scope of our paper and such detailed studies including  numerical simulations on  DM accretion along with  consequent constraints on the PBH/DM parameters are left for future work.
Light DM is a relatively new avenue  and indeed less explored compared with heavier DM ($>1$ GeV). We briefly discuss these implications, before concluding our discussion with observational prospects based on our boost factor estimates for our PBH isocurvature scenarios.

Even though  WIMPs from thermal freeze-out are indeed well motivated from  theory, facing the lack of any observational data hints of   canonical properties of the thermal WIMP DM,  growing attention has been paid to  beyond-WIMP possibilities such as light (sub-GeV) DM \cite{Knapen:2017xzo,Battaglieri:2017aum,Bondarenko:2019vrb,Coogan:2019qpu,Essig:2013goa,Cirelli:2020bpc,Choudhury:2019tss,Hooper:2007tu,Tashiro:2014tsa,Xu:2018efh,Ooba:2019erm}.
Light DM search experiments include  X-ray and gamma-ray telescopes sensitive to  sub-GeV DM annihilations/decays such as EGRET, COMPTEL, INTEGRAL, HEAO-1 as well as the planned e-ASTROGRAM, AMEGO and eROSITA \cite{Strong:2004de,comptel,1993ApJS...86..629T,Gruber:1999yr,DeAngelis:2017gra,McEnery:2019tcm, Merloni:2012uf}. Consequently there have been studies on indirect searches for sub-GeV DM annihilation/decay for a wide range of DM scenarios \cite{Essig:2013goa,Cirelli:2020bpc,Leane:2020liq,Boddy:2015efa,Bartels:2017dpb}.
For instance, for  decaying DM whose mass is of order a few keV to a few GeV, the lower bounds on the DM lifetime $\tau$ are in the range of $10^{24}\sim 10^{28}$ sec depending on whether the produced photons are from  direct decays  or  final state radiation \cite{Essig:2013goa}.
We can simply translate these bounds to gain insight on the potential significance of the PBHs as follows.   
Because the subhalos follow the underlying DM density distributions (rather than the DM density squared), our unresolved diffuse gamma ray backgrounds observationally resemble those of decaying DM rather than the conventional annihilation signals \cite{Yang:2011eg,Boucenna:2017ghj,Blanco:2019eij,Delos:2019dyh}.
We can hence, for a simple order-of-magnitude estimate, equate the total annihilation rate to the total decay rate for a given region of interest
\ba
\frac{\langle \sigma v \rangle}{2 m_{\chi}^2} \sum_i  \int d^3 x \rho^2_{i}
\approx 
\frac{1}{\tau} \frac{\int d^3 x \rho_{mat}}{m_{\chi}}
\ea
where the sum of the subhalo $J$ factors is given by Eq. (\ref{shboost}). 
For instance, for $m_{\chi}=1$ MeV, the bound $\tau \gtrsim 10^{26}$ sec can be translated into the annihilation bound of order $\langle \sigma v \rangle \lesssim 10^{-32} cm^3/s$ for $f_{PBH}=0.01, M_{PBH}=0.1 M_{\odot}$. This changes to $\langle \sigma v \rangle \lesssim 10^{-31} cm^3/s$ for $f_{PBH}=0.01, M_{PBH}=10^{-6} M_{\odot}$  according to the changes in the boost factors shown in Fig. \ref{boostvdmhost}. 
These bounds can be tighter than the current CMB bounds ($\langle \sigma v\rangle \lesssim 10^{-30} cm^3/s$ for  MeV mass DM with  s-wave annihilations) which limits the annihilation cross section to be below the canonical thermal WIMP value \cite{Slatyer:2015jla,Essig:2013goa}. A more precise estimate should take account of the experimental specifications and the particle physics model details such as the exact decay channels, but our estimate allows an insight into the potential significance of the fractional PBH DM for structure formation and the motivations for investigating further details such as more detailed subhalo survival probabilities in the mixed DM scenarios.
\\
\\
We have studied the effects of PBH Poisson fluctuations to enhance the small scale structures, and the DM distributions at the present epoch can be consequently clumpier than the conventional $\Lambda$CDM scenario. As an interesting phenomenology for applying such effects, we discussed  DM annihilation boosts due to enhanced substructures. Such early-formed minihalos should be also of great interest for other observations such as  21cm signals \cite{Cole:2019zhu,Morabito:2020glh,shi2019,Kadota:2020ybe,Irsic:2019iff,Hutsi:2019hlw}. Dedicated numerical studies are warranted to study minihalo evolution in mixed DM scenarios. Lacking  evidence of  DM in the conventional WIMP parameter range,  DM studies have accordingly  been diversified. Interesting possible DM candidates include light (sub-GeV) DM as well as PBHs which gained growing interest in connection with the advancement of gravitational wave astronomy. The co-existence of  annihilating DM and  PBHs is an intriguing possibility which could well unveil unexpected DM properties through detection in indirect DM search experiments. We treated the dark matter particle mass and  interaction rates as free parameters without specifying their origins, and a more model-dependent analysis with the motivated particle physics models would be also worth pursuing.
\\
\\
This work was supported by the Institute for Basic Science (IBS-R018-D1) and IAP/Sorbonne  Universit\'{e}. 





\newpage
\bibliography{./kenjireference}



\end{document}